\documentclass{article}
\usepackage{fullpage}
\usepackage{amsmath,amsthm,amssymb, color, tikz, mathtools}
\pdfoutput=1

\usetikzlibrary{positioning}
\usetikzlibrary{shapes}
\usetikzlibrary{decorations.pathreplacing}
\usepackage{pgf-pie}

\usepackage{tcolorbox}

\usepackage{enumerate, enumitem} 
\usepackage{hyperref} 
\hypersetup{colorlinks,linkcolor={blue},citecolor={blue},urlcolor={red}}

\usepackage{graphicx}
\usepackage[normalem]{ulem}

\newcommand{\braket}[2]{\left \langle #1 | #2 \right\rangle}
\newcommand{\ket}[1]{|#1\rangle}

\definecolor{Qorange1}{rgb}{1.0, 0.27, 0.0}
\definecolor{Qorange2}{rgb}{1.0, 0.63, 0.54}
\definecolor{Qblue2}{RGB}{46, 109, 189}
\definecolor{Qblue1}{rgb}{0.67, 0.8, 0.94}

\definecolor{Qblue2M}{RGB}{214,222,231}
\definecolor{Qorange2M}{RGB}{237,229,225}

\newcommand{\smalldiscsize}{0.8}
\newcommand{\discsize}{1.0}
\newcommand{\innerdiscsize}{0.8}
\newcommand{\outerdiscsize}{1.2}

\tikzset{
     lines/.style={draw=none},
}

\usepackage{intcalc}
\newcommand{\sign}[1]{
	\ifcase\intcalcSgn{#1}~\or~\else-\fi
}

\newcommand{\qubobLabel}[2]{ 
}

\newcommand\qbox[1]{\textcolor{#1}{\rule{3mm}{3mm}}}

\makeatletter
\newcounter{c}
\newcommand{\select}[2]{
\setcounter{c}{1}
\@for\x:={#1}\do{\ifnum \value{c}=#2 {\x} \else{ }\fi \stepcounter{c}}
}
\makeatother

\newcounter{qketOne}
\renewcommand*{\theqketOne}{
\ifnum\value{qketOne}=0{$\ket{0}$}\fi
\ifnum\value{qketOne}=1{$\ket{1}$}\fi
\ifnum\value{qketOne}=2{$\ket{0}$}\fi
\ifnum\value{qketOne}=3{$\ket{1}$}\fi
\ifnum\value{qketOne}=4{$\ket{0}$}\fi
\ifnum\value{qketOne}=5{$\ket{1}$}\fi
\ifnum\value{qketOne}=6{$\ket{0}$}\fi
\ifnum\value{qketOne}=7{$\ket{1}$}\fi
}
\newcounter{qlabelOne}
\renewcommand*{\theqlabelOne}{
\ifnum\value{qlabelOne}=0{\qbox{Qblue1}}\fi
\ifnum\value{qlabelOne}=1{\qbox{Qorange1}}\fi
\ifnum\value{qlabelOne}=2{\qbox{Qblue1}}\fi
\ifnum\value{qlabelOne}=3{\qbox{Qorange1}}\fi
\ifnum\value{qlabelOne}=4{\qbox{Qblue1}}\fi
\ifnum\value{qlabelOne}=5{\qbox{Qorange1}}\fi
\ifnum\value{qlabelOne}=6{\qbox{Qblue1}}\fi
\ifnum\value{qlabelOne}=7{\qbox{Qorange1}}\fi
}

\newcounter{qlabelTwo}
\renewcommand*{\theqlabelTwo}{
\ifnum\value{qlabelTwo}=0{\qbox{Qblue2}}\fi
\ifnum\value{qlabelTwo}=1{\qbox{Qorange2}}\fi
\ifnum\value{qlabelTwo}=2{\qbox{Qblue2}}\fi
\ifnum\value{qlabelTwo}=3{\qbox{Qorange2}}\fi
\ifnum\value{qlabelTwo}=4{\qbox{Qblue2}}\fi
\ifnum\value{qlabelTwo}=5{\qbox{Qorange2}}\fi
\ifnum\value{qlabelTwo}=6{\qbox{Qblue2}}\fi
\ifnum\value{qlabelTwo}=7{\qbox{Qorange2}}\fi
}

\newcounter{twoqket}
\renewcommand*{\thetwoqket}{
\ifnum\value{twoqket}=0{$\ket{00}$}\fi
\ifnum\value{twoqket}=1{$\ket{01}$}\fi
\ifnum\value{twoqket}=2{$\ket{11}$}\fi
\ifnum\value{twoqket}=3{$\ket{10}$}\fi
\ifnum\value{twoqket}=4{$\ket{00}$}\fi
\ifnum\value{twoqket}=5{$\ket{01}$}\fi
\ifnum\value{twoqket}=6{$\ket{11}$}\fi
\ifnum\value{twoqket}=7{$\ket{10}$}\fi
}
\newcounter{twoqlabel}
\renewcommand*{\thetwoqlabel}{
\ifnum\value{twoqlabel}=0{\qbox{Qblue1}\qbox{Qblue2}}\fi
\ifnum\value{twoqlabel}=1{\qbox{Qblue1}\qbox{Qorange2}}\fi
\ifnum\value{twoqlabel}=2{\qbox{Qorange1}\qbox{Qorange2}}\fi
\ifnum\value{twoqlabel}=3{\qbox{Qorange1}\qbox{Qblue2}}\fi
\ifnum\value{twoqlabel}=4{\qbox{Qblue1}\qbox{Qblue2}}\fi
\ifnum\value{twoqlabel}=5{\qbox{Qblue1}\qbox{Qorange2}}\fi
\ifnum\value{twoqlabel}=6{\qbox{Qorange1}\qbox{Qorange2}}\fi
\ifnum\value{twoqlabel}=7{\qbox{Qorange1}\qbox{Qblue2}}\fi
}

\newcounter{n}
\makeatletter
\newcommand{\newlegend}[1]{
\parbox[b]{2cm}{
\setcounter{twoqlabel}{0}
\setcounter{twoqket}{0}
\setcounter{n}{1}
\@for\slice:={#1}\do{
	\thetwoqlabel~\slice\thetwoqket \\
	\stepcounter{twoqlabel}
	\stepcounter{twoqket}
	\stepcounter{n}
}
\ifnum\value{n}>8 \vspace{-6ex} \else \vspace{-1ex}\fi
}
\ifnum\value{n}>8 \vspace{2ex} \else \fi
}

\newcommand{\newlegendOne}[1]{
\parbox[b]{2cm}{
\setcounter{qlabelOne}{0}
\setcounter{qketOne}{0}
\setcounter{n}{1}
\@for\slice:={#1}\do{
	\theqlabelOne~\slice\theqketOne \\
	\stepcounter{qlabelOne}
	\stepcounter{qketOne}
	\stepcounter{n}
}
\ifnum\value{n}>8 \vspace{-6ex} \else \vspace{-1ex}\fi
}
}
\newcommand{\newlegendTwo}[1]{
\parbox[b]{2cm}{
\setcounter{qlabelTwo}{0}
\setcounter{qketOne}{0}
\setcounter{n}{1}
\@for\slice:={#1}\do{
	\theqlabelTwo~\slice\theqketOne \\
	\stepcounter{qlabelTwo}
	\stepcounter{qketOne}
	\stepcounter{n}
}
\ifnum\value{n}>8 \vspace{-6ex} \else \vspace{-1ex}\fi
}
}
\makeatother

\newcommand{\qubobValue}[1]{ \intcalcAbs{#1}}

\newcommand{\qubobPie}[5]{
  \pie[ text=inside, 
	hide number,
	style={lines}, change direction,
	rotate=#1,
	pos={#2}, 
	radius=#3,
 	color={#4}]
	{ #5 }
}

\newcommand{\qubobOne}[2]{
\begin{tikzpicture}
  \qubobPie{90}{8,0}{\discsize}{Qblue1,Qorange1}
	{ \qubobValue{#1}/\qubobLabel{#1}{0},
	  \qubobValue{#2}/\qubobLabel{#2}{1}
	 }
\end{tikzpicture}
\newlegendOne{\sign{#1},\sign{#2}}
}

\newcommand{\qubobOneSplit}[4]{
\begin{tikzpicture}
  \qubobPie{90}{8,0}{\discsize}{Qblue1,Qorange1,Qblue1, Qorange1}
	{ \qubobValue{#1}/ \qubobLabel{#1}{0}, 
	  \qubobValue{#2}/ \qubobLabel{#2}{1},
	  \qubobValue{#3}/  \qubobLabel{#3}{0},
          \qubobValue{#4}/  \qubobLabel{#4}{1}
	}
\end{tikzpicture}
\newlegendOne{\sign{#1},\sign{#2},\sign{#3},\sign{#4}}
}

\newcommand{\qubobsSide}[4]{
\begin{tikzpicture}
  \qubobPie{90}{0,0}{\discsize}{Qblue1,Qorange1}
	{ #1+#2/\qubobLabel{}{},
	#4+#3/\qubobLabel{}{}}
  \hspace {-1cm}\qubobPie{90+3.6*(#4)}{5,0}{\discsize}{Qblue2,Qorange2}
	{ #1+#4/\qubobLabel{}{},
	#3+#2/\qubobLabel{}{}}
\end{tikzpicture}
}

\newcommand{\qubobsStacked}[4]{
  \begin{tikzpicture}
  \qubobPie{90}{0,0}{\outerdiscsize}{Qblue1,Qorange1 }
	  {#1+#2/\qubobLabel{}{},
	#3+#4/\qubobLabel{}{}}
  \qubobPie{90+3.6*#4}{0,0}{\innerdiscsize}{Qblue2,Qorange2 }
	  { #1+#4/\qubobLabel{}{},
	#3+#2/\qubobLabel{}{}}
  \end{tikzpicture}
  \newlegend{\sign{#1},\sign{#2},\sign{#3},\sign{#4}}
}

\newcommand{\qubobsTensorStacked}[4]{
  \qubobsStacked{#1*#3*0.01}{#1*#4*0.01}{#2*#4*0.01}{#2*#3*0.01}
}

\newcommand{\measureZero}[4]{
\begin{tikzpicture}
  \qubobPie{90}{0,0}{\outerdiscsize}{Qblue1,lightgray}
	{ #1+#2/\qubobLabel{}{},
	#4+#3/\qubobLabel{}{} }
  \qubobPie{90}{0,0}{\innerdiscsize}{Qblue2,Qorange2,Qorange2M,Qblue2M}
	{  #1/\qubobLabel{}{}, 
	#2/\qubobLabel{}{}, 
	#3/\qubobLabel{}{},
	#4/\qubobLabel{}{} }
\end{tikzpicture}
}

\newcommand{\measureOne}[4]{
\begin{tikzpicture}
  \qubobPie{90}{0,0}{\outerdiscsize}{lightgray,Qorange1}
	{ #1+#2/\qubobLabel{}{}, #4+#3/\qubobLabel{}{}}
  \qubobPie{90}{0,0}{\innerdiscsize}{Qblue2M,Qorange2M,Qorange2,Qblue2}
	{ #1/\qubobLabel{}{},#2/\qubobLabel{}{},#3/\qubobLabel{}{},
	#4/\qubobLabel{}{}}
\end{tikzpicture}
}

\newcommand{\half}[1]{\intcalcDiv{#1}{2}}

\newcommand{\teleportationSetup}[2]{
\qubobsStacked{\half{#1}}{\half{#2}}{\half{#2}}{\half{#1}}
\qubobOne{50}{50}
}

\newcommand{\teleportationCNOT}[2]{
\begin{tikzpicture}
	\qubobPie{90}{0,0}{\outerdiscsize}{Qblue1,Qorange1}
	{ \half{#1}/ , \half{#2}/ , \half{#2}/ ,\half{#1}/ }
	\qubobPie{90+#1/2*3.6}{0,0}{\innerdiscsize}{Qblue2,Qorange2 }
	{#1/ ,  #2/ }
\end{tikzpicture}
\newlegend{~,~,~,~}
\qubobOne{50}{50}
}

\newcommand{\teleportationMeasureZero}[2]{
\begin{tikzpicture}
\qubobPie{90}{0,0}{\outerdiscsize}{Qblue1,lightgray,Qblue1,lightgray}
	{ \half{#1}/ , \half{#2}/ , \half{#2}/ ,\half{#1}/ }
\qubobPie{90}{0,0}{\innerdiscsize}{Qblue2, lightgray, Qorange2, lightgray}
	{ \half{#1}/ , \half{#2}/ , \half{#2}/ ,\half{#1}/ }
\end{tikzpicture}
\newlegend{~,~,~,~}
\begin{tikzpicture}
\qubobPie{90}{0,0}{\discsize}{Qblue1,lightgray,Qorange1,lightgray}
	{ \half{#1}/ , \half{#2}/ , \half{#2}/ ,\half{#1}/ }
\end{tikzpicture}
\newlegendOne{~,~}
}

\newcommand{\teleportationMeasureOne}[2]{
\begin{tikzpicture}
\qubobPie{90}{0,0}{\outerdiscsize}{lightgray,Qorange1,lightgray,Qorange1}
	{\half{#1}/ , \half{#2}/ , \half{#2}/ ,\half{#1}/ }
\qubobPie{90}{0,0}{\innerdiscsize}{lightgray, Qorange2, lightgray, Qblue2}
	{ \half{#1}/ , \half{#2}/ , \half{#2}/ ,\half{#1}/ }
\end{tikzpicture}
\newlegend{~,~,~,~}
\begin{tikzpicture}
\qubobPie{90}{0,0}{\discsize}{lightgray,Qblue1,lightgray,Qorange1}
	{ \half{#1}/ , \half{#2}/ , \half{#2}/ ,\half{#1}/ }
\end{tikzpicture}
\newlegendOne{~,~}
}

\newcommand{\teleportationH}[2]{
\begin{tikzpicture}
\qubobPie{90}{0,0}{\outerdiscsize}{Qblue1,Qorange1,Qblue1,Qorange1}
	{ \half{#1}/ , \half{#2}/ , \half{#2}/ ,\half{#1}/ }
\qubobPie{90}{0,0}{\innerdiscsize}{Qblue2, Qorange2, Qblue2, Qorange2}
	{ \half{#1/2}/ , \half{#1/2}/ , \half{#2/2}/ , \half{#2/2}/ ,
	  \half{#2/2}/ , \half{#2/2}/ ,\half{#1/2}/ , \half{#1/2}/ }
\end{tikzpicture}
\newlegend{~,~,-,~,~,-,~,~}
\begin{tikzpicture}
\qubobPie{90}{0,0}{\discsize}{Qblue1,Qorange1}
	{ 50/ , 50/ }
\end{tikzpicture}
\newlegendOne{~,~}
}

\newcommand{\teleportationMeasureZZ}[2]{
\begin{tikzpicture}
\qubobPie{90}{0,0}{\smalldiscsize}{Qblue1,lightgray,Qorange1,lightgray}
	{ \half{#1/2}/ , \half{#1/2}+\half{#2/2}+\half{#2/2}/ ,
	  \half{#2/2}/ , \half{#2/2}+\half{#1/2}+\half{#1/2}/ }
\end{tikzpicture}
\newlegendOne{~,~}
}
\newcommand{\teleportationMeasureZO}[2]{
\begin{tikzpicture}
\qubobPie{90}{0,0}{\smalldiscsize}{lightgray,Qblue1,lightgray,Qorange1,lightgray}
	{ \half{#1/2}/ , \half{#1/2}/ , \half{#2/2}+\half{#2/2}+\half{#2/2}/ ,
	  \half{#2/2}/ ,\half{#1/2}+\half{#1/2}/ }
\end{tikzpicture}
\newlegendOne{~,-}
}

\newcommand{\teleportationMeasureOZ}[2]{
\begin{tikzpicture}
\qubobPie{90}{0,0}{\smalldiscsize}{lightgray,Qblue1,lightgray,Qorange1,lightgray}
	{ \half{#1/2}+\half{#1/2}/ ,  \half{#2/2}/ , 3*\half{#2/2}/ ,
	  \half{#1/2}/ , \half{#1/2}/  }
\end{tikzpicture}
\newlegendOne{~,~}
}

\newcommand{\teleportationMeasureOO}[2]{
\begin{tikzpicture}
\qubobPie{90}{0,0}{\smalldiscsize}{lightgray,Qblue1,lightgray,Qorange1}
	{ \half{#1/2}+\half{#1/2}+\half{#2/2}/ , \half{#2/2}/ ,
	  \half{#2/2}+\half{#2/2}+\half{#1/2}/ , \half{#1/2}/ }
\end{tikzpicture}
\newlegendOne{~,-}
}

\makeatother

\newcommand{\mathify}[1]{\ifmmode{#1}\else\mbox{$#1$}\fi}

\newcommand{\sgn}{\mathrm{sgn}}

\newcommand{\Reals}{\mathbb{R}}

\usepackage{tcolorbox}

\usepackage{bbm} 
\usepackage{mathtools}

\newcommand{\hide}[1]{}

\title{QuBOBS, interactive objects and a visual representation\\
to explain quantum computing}
\author{Sophie Laplante\thanks{IRIF, Universit\'e Paris Cit\'e} \and Loris Perez\thanks{Musician and composer} \and Sylvie Tissot\thanks{Software engineer, Anabole} \and Lou Vettier\thanks{Designer, {\tt louvettier.com}}}

\date{}

\begin{document}
\maketitle

\begin{abstract}
We introduce a visual representation of qubits to assist in explaining quantum 
computing to a broad audience.
The representation follows from physical devices that we developed to
explain superposition, entanglement, measurement, phases, interference, 
and quantum gates.
We describe how this representation can be used to explain teleportation
and the Bennett-Brassard quantum key agreement protocol.
\end{abstract}

\section{Introduction}

This project is the result of a collaboration between a computer scientist, a designer, a software developer and a composer. Together, we are developing tools to make quantum computing accessible to a wider audience, without sacrificing (too much) mathematical correctness. 
In this paper, we describe the mathematics behind the objects that we are developing.
We assume that the reader is familiar with the basics of quantum computing. Our goal is to 
provide the reader with a representation that can be used for teaching at the undergraduate and graduate level, to students with or without mathematical background, as well as for outreach activities.

Most researchers in quantum information and algorithms agree that
explaining quantum computing to a wide audience without relying on
mathematical machinery is a difficult task. A recent article of Aaronson lays out the
difficulties particularly well~\cite{Aaronson21}.
Many different approaches have been used to explain quantum computing to a wide audience, and although there are far too many to give an exhaustive survey here, we give a few examples. 

For single qubit applications such as key agreement, Charles Bennett~\cite{Bennett} 
and many others use double arrows to represent photon polarization. 
John Preskill~\cite{Preskill}  uses  colored balls to represent 0 and 1. 
Qubits are represented as boxes and gates with two doors. Putting in a qubit in
the top door and taking it out of the bottom door applies a change of basis.
This representation is used for single qubit protocols as well as to explain 
entanglement.

Karl Svovil explains quantum cryptography using chocolate balls~\cite{Svovil}
with a 0 or 1 symbol on each ball, in either green or in red to indicate the basis. 
Special glasses (red and green) are used to explain measurement in the correct basis.

The most complete representation we are aware of 
is proposed by Terry Rudolph and coauthors~\cite{Rudolph, Rudolph05,Economou}. This
representation is intended to explain quantum circuits and computation in full generality, without using mathematical symbols.
Qubits are presented as clouds of black and white balls, in a proportion that 
corresponds to the probability of getting the corresponding outcome when 
measuring the qubit.
Blocks represent operations on qubits. Using this representation, they explain
several advanced applications, such as Grover search~\cite{G1996} and CHSH games~\cite{CHSH}.

For more advanced topics in quantum information, tensor networks have
been used as a graphical language to represent quantum states and processes~\cite{Wood15, Biamonte19}.

Our approach has been to construct simple physical devices to help the 
audience develop a very concrete mental image of concepts that
can be difficult to grasp, such as randomness which is inherent to
quantum computing, and entanglement. Special care has gone towards distinguishing randomness and truly
quantum properties. 
Superposition is represented by two overlapping paper disks.
Measurement is represented by spinning a window over a disk and observing the result 
through a small window. Correlation is made concrete by having two windows spinning
together as cog wheels.
Once the physical representation is clear, we move on to a slightly more 
abstract graphical representation, on which we can apply various operations.

\section{The QUBOBS representation }
\subsection{Qubits}

Our main contribution is a simple, robust and surprisingly versatile representation of qubits. A qubit 
$\alpha \ket{0} +\beta\ket {1}$ is represented by a two-color disc. Blue represents $\ket{0}$ and
orange represents $\ket{1}$. The proportion of each color represents the probability of measuring the 
corresponding outcome, hence the blue part takes up an $|\alpha|^2$ fraction of the disc's area 
(and circumference) and the orange part takes up $|\beta|^2$.

Most elementary quantum algorithms use signed reals as amplitudes so to complete the description, we use a negative sign to indicate when the phase is -1.

Our qubit representation is made of colored paper discs. Two discs can be interleaved 
to obtain superpositions with arbitrary amplitudes.

\begin{figure}
\centering
\includegraphics[height=3cm]{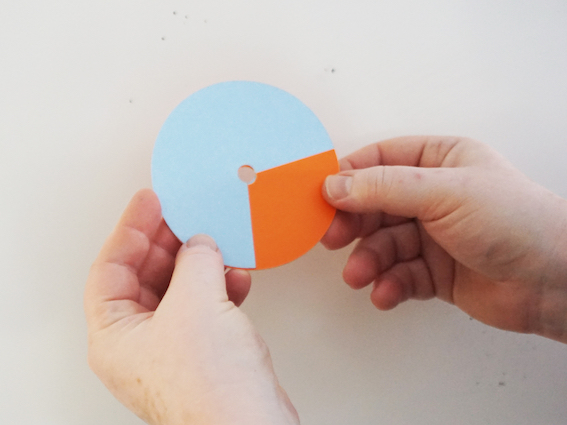}
\includegraphics[height=3cm]{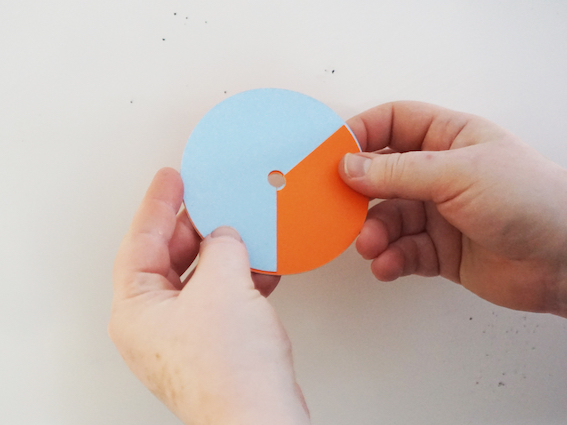}
\centering
\includegraphics[height=3cm]{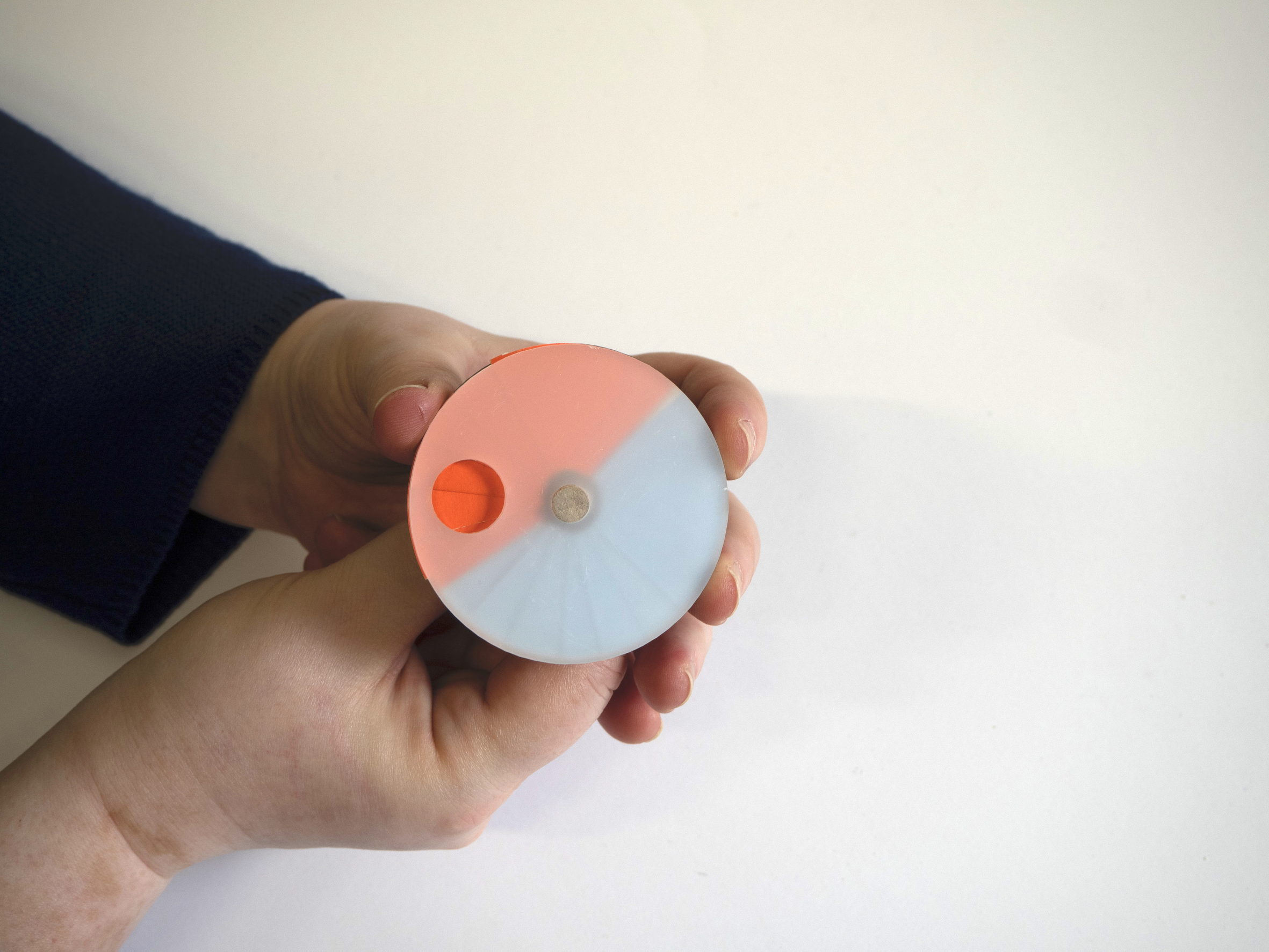}
\caption{
An object that represents a quantum state $\ket{\phi}$. The squared magnitude of $\braket{\phi}{0}$ is blue,
and the squared magnitude of $\braket{\phi}{1}$ is orange. On the device on the right, the window spins around the edge of the qubit, When it stops, the color under the window is the outcome of the measurement.}
\end{figure}

We illustrate measurements with a small device 
that spins around the qubit. When it stops, a small round window 
reveals one of the colors. This is the outcome of the measurement. The probability of each outcome 
is the part of the disc that is the corresponding color, which is exactly the probability of
measuring the qubit in the standard basis. 

Taking the modulus square of the amplitude sacrifices the phase, but we find that this is a useful first step
towards understanding the inherent randomness of quantum measurements. When we use our devices, we 
show how this 
is also how we could represent a biased coin.  Flipping a coin and observing the outcome is similar
to measuring a qubit, but as we point out, a coin flip is not a quantum phenomenon.
This allows us to 
highlight the fact that the phase is a crucial part of a quantum state, and that without phases, 
we are only describing probabilistic binary states, and probabilistic computation.

Most elementary quantum algorithms use only positive and negative reals and do not 
require the complex numbers in their full generality. We introduce the phase as a sign, 
which is sufficient for quantum computing.

We write $\ket{\phi}\mapsto (P,Q)$ to mean that $(P,Q)$ is our representation of $\ket{\phi}$.  
Using this notation, a qubit $\alpha \ket{0} +\beta\ket {1}$ with $\alpha,\beta\in \Reals$
is represented by $(sgn(\alpha)|\alpha|^2, sgn(\beta) |\beta|^2)$, and conversely, $(A,B)$ is a 
representation of the qubit $sgn(A)\sqrt{|A|}\ket{0} + sgn(B)\sqrt{|B|}\ket{1}$.
When several qubits are involved, the angle of the disk becomes important, and we will assume that, unless indicated otherwise, the blue part starts at the top going clockwise.

\subsection{Entanglement and  partial measurements}

The surprising expressiveness of our representation of qubits appears when we consider two qubits.
Separable states can be represented by two observation windows spinning independently over two qubits.
However, things become interesting when we link the two spinning windows with a cog. so that they spin 
synchronously. This allows us to represent entanglement as we explain now.

\subsubsection{Entanglement}

\begin{figure}[h]
\centering
\includegraphics[height=3cm]{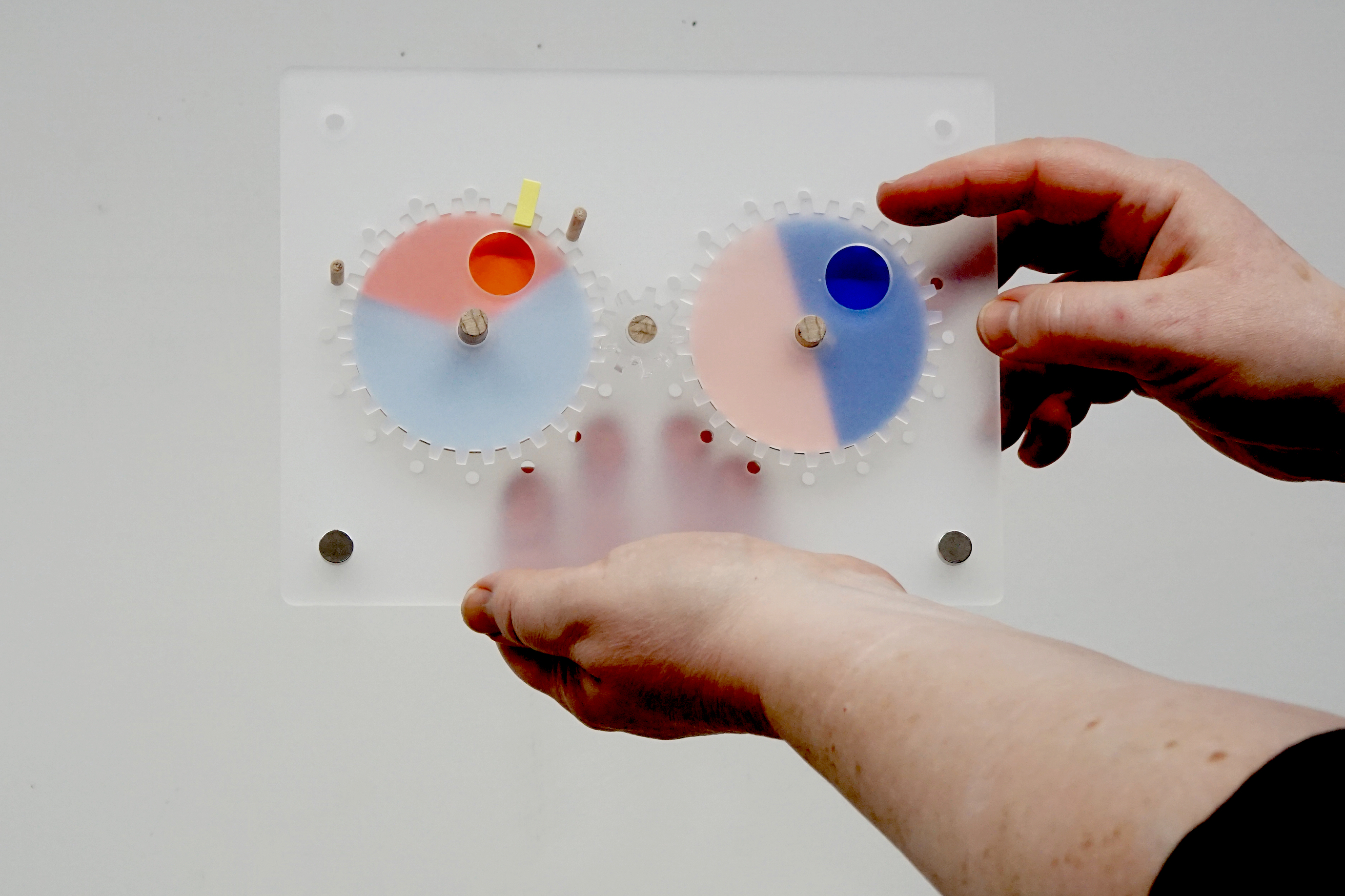}
\includegraphics[height=3cm]{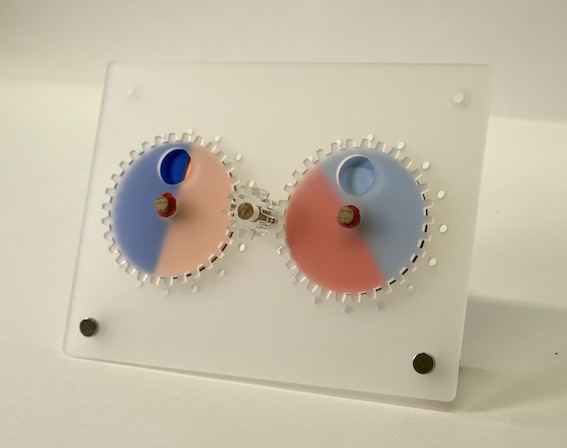}
\includegraphics[height=3cm]{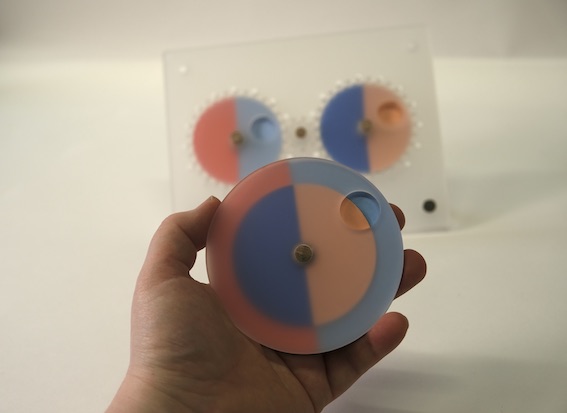}
\caption{Two ways of representing two entangled qubits. On the left and in the middle, a cog makes both windows spin synchronously. On the right, the qubits are stacked and a single window lets us observe both qubits.  The state represented on the right on both devices is $\frac{1}{\sqrt{2}}(\ket{01}+\ket{10})$.}
\label{fig:ent}
\end{figure}

Consider an arbitrary two-qubit pure state 
$\ket{\phi} = \alpha \ket{00} + \beta\ket{01}+ \gamma\ket{11} + \delta\ket{10}$.
(Notice that we have used a Gray code to order the basis elements.)
If the first qubit is measured with a standard measurement, the outcome 
is 0 with probability $|\alpha|^2 + |\beta|^2$ and 1
with probability  $|\gamma|^2 + |\delta|^2$. Similarly, if the second qubit is measured, the outcome
is $0$ with probability $|\alpha|^2+|\delta|^2$ and 1 with probability $|\beta|^2+|\gamma|^2$.
This allows us to represent the two-qubit state with one disc that has  $|\alpha|^2 + |\beta|^2$ fraction of blue and  $|\gamma|^2 + |\delta|^2$ of orange, and the second disc 
with  $|\alpha|^2+|\delta|^2$ blue and  $|\beta|^2+|\gamma|^2$ orange. They key observation is that when the observation windows are lined up (say they both start from the top of the disk and turn synchronously)
it is always possible to line up the two discs so that the probability of getting 00 when 
both qubits are measured is $|\alpha|^2$, and similarly for 01, 11, and 10, as illustrated below.
To do this, line up the first qubit so that it is blue starting from the top going clockwise, and line up
the second qubit so that it is orange starting at an angle $\theta= |\alpha|^2 \cdot 2\pi$ from the top,
going clockwise. Then starting from the top and going clockwise, the two qubits are blue for 
$|\alpha|^2 \cdot 2\pi$, then the first qubit is blue and the second is orange for $|\beta|^2\cdot 2\pi$ and so on.

We use a second device to visualize the four areas corresponding to each of the outcome pairs 00,01, 11, and 10 (Figure~\ref{fig:ent}, on the right). We stack the two qubits on top of each other, with a single window going around the perimeter. Going back and forth between these two representations, side by side and stacked,
gives us two complementary ways of seeing the effects of entanglement.

\begin{figure}
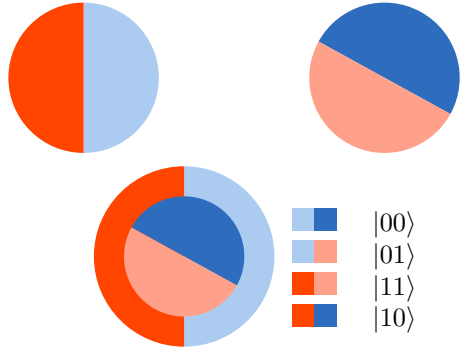

\centering
\qubobsSide{33}{17}{33}{17} \\
\qubobsStacked{33}{17}{33}{17}
\caption{Two representations of the same pair of entangled qubits, $\frac{1}{\sqrt{3}}\ket{00}+\frac{1}{\sqrt{6}}\ket{01}+\frac{1}{\sqrt{3}}\ket{11}+\frac{1}{\sqrt{6}}\ket{10}$. On top they are represented side by side like
in the device with
two windows spinning synchronously. On the bottom the qubits are stacked the same way as in the device 
with a single window.}
\label{fig:diagrams}
\end{figure}

When explaining protocols or algorithms, we set aside the physical devices with the windows spinning around the 
perimeter, and work with the diagrams illustrated in Figure~\ref{fig:diagrams}.

Separable states can also be represented with synchronized observation windows.
For example, the state $\frac {1} {\sqrt{2}}( \ket{0}+\ket{1} ) \otimes \frac {1} {\sqrt{2}}( \ket{0}+\ket{1} ) $
can be represented with two half blue, half orange disks placed perpendicularly.

\begin{figure}
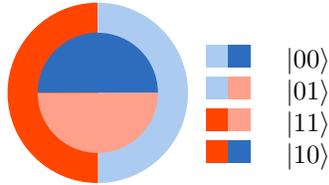

\centering\qubobsTensorStacked{50}{50}{50}{50}
\caption{ Our representation of the separable state $\frac{1}{\sqrt{2}}(\ket{0}+\ket{1})\otimes \frac{1}{\sqrt{2}}(\ket{0}+\ket{1})$}
\end{figure}

This works well for two qubits. From three qubits onward we can no longer represent arbitrary
states with discs with two colors arranged into two contiguous colored slices. However, 
it turns out that for many applications, two qubits suffice, and furthermore, 
presenting qubits using more than two color slices occurs naturally as a result of applying
quantum gates.

\subsubsection{Partial measurements}

Partial measurements of a bipartite state and the residual state after a measurement is made
is generally viewed as difficult to explain without appealing to a formal projection operator.
Our representation is surprisingly successful in conveying the math without writing 
out any equations.

\begin{figure}
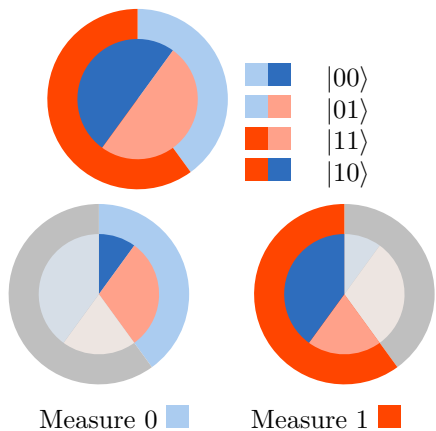
\centering
\qubobsStacked{10}{30}{20}{40}
\\
\centering
\measureZero{10}{30}{20}{40} \quad
\measureOne{10}{30}{20}{40}

Measure 0 \qbox{Qblue1}  \quad \quad Measure 1 \qbox{Qorange1}

\caption{Visualizing the effect of a partial measurement. On the top, the qubit before measuring the fist (outer) qubit.
Below, the residual state when the outcome is blue (on the left) or orange (on the right).}
\label{fig:measure}
\end{figure}

Returning to our two-qubit state 
$\ket{\phi} = \alpha \ket{00} + \beta\ket{01}+ \gamma\ket{11} + \delta\ket{10}$, 
let us make a measurement on the first qubit. The blue and orange parts of the first qubit 
split the second qubits into two areas. In Figure~\ref{fig:measure} we give an example of how we can visualize 
the residual state of the second qubit given that the first qubit has been measured. 
If the first qubit came out blue, then the observation window could have fallen in any place where the first qubit was blue. This part of the disk, on both qubits, since the measurement windows are linked by a cog, is the residual state.
The rest of the disks are no longer accessible.
Having this visual representation in mind makes it easier for students to conceptualize what the projection does 
and what the renormalization of the state accomplishes (not to mention what the norm actually represents).

\subsection{Phases and interference}

In our basic representation, we restrict ourselves to +/-1 phases.
In the simpler quantum applications, such as key exchange or Deutsch's and Deutsch-Jozsa's algorithm,
when interference occurs, opposing amplitudes cancel exactly.
By this we mean that in the course of the computation, we obtain states of the form
$\alpha \ket{0} - \alpha' \ket{0} + \beta\ket{1} + \beta' \ket{1}$  where $\alpha=\alpha'$.
This is a great stroke of luck since  $|\alpha - \alpha'|^2=|\alpha|^2 - |\alpha'|^2$ = 0, and  the slices with opposing phases cancel exactly. Obviously this does not hold in general.\footnote{The incorrect identity, $(x+y)^2 = x^2+y^2$ has been coined ``freshman's dream'' 
following Saunders Mac Lane~\cite{MacLane} who attributes it to Stephen Kleene.}
However, the fact that this happens to be correct in some key cases of interest 
allows us  to convey how interference plays
a role in quantum computation, in a meaningful and (almost) mathematically correct way.
To more mathematically advanced audiences, we always point out that we are taking a sinificant shortcut here,
and in a classroom situation we can return to the more conventional mathematical representation,
and address other instances  of interference later on, when discussing Grover's algorithm, 
or quantum strategies for Bell inequality violations for instance.

\begin{figure}[h]
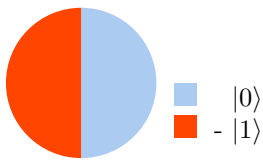

\centering\qubobOne{50}{-50}
\caption{The qubit $\frac{1}{\sqrt{2}}(\ket{0}-\ket{1})$ represented as $(1/2,-1/2)$.}
\end{figure}

When explaining interference, we have found that referring to colors (blue, orange) 
makes it easier to grasp what is meant by interference than when using numerical values 0 and 1.
We have often been asked what it means for +0 and -0 to cancel, since $+0-0=0$. This problem of 
confusing the state with its label does not occur when talking about colors.
It also offers an opportunity to discuss the fact that blue and orange can be labels for many
different propoerties, such as spin, or polarization, as long as these properties are fully 
distinguishible.

To help explain phases and interference, we appeal to the audience's knowledge that 
light behaves like a wave. Positive and negative phases can be 
illustrated with sine waves that can be shifted by half a wavelength.%
\footnote{For more advanced audiences, one could generalize to arbitrary shifts between 0 and half of
the wavelength but we have not found this to be necessary to explain basic quantum algorithms and 
protocols.}
For younger audiences, we have illustrated this with two people pulling and pushing 
on two ends of a table in a synchronized cyclical motion. If both have the same phase, one 
pushing as the other pulls, then the amplitudes add up. If they have opposite phases, they cancel 
each other and the table doesn't move. Pushing on a blue table does not interfere with pushing on 
an orange table.
Many audience members are familiar with this phenomenon in the realm of sound waves and 
have pointed out that this is how noise-canceling headphones work.

\subsection{Single qubit gates and linearity}

We illustrate quantum gates by flipping over a disk to reveal the effect of a gate on each of the basis states.\footnote{Many of the gates we consider are self-adjoint, and flipping over twice gets us back to the initial state.}

In order to explain simple quantum circuits, it would be nice to apply quantum gates on our representation of qubits by 
appealing to linearity. This works well in some cases, like the X gate, or the Z gate.

However, 
we have to proceed with caution with other gates, such as the Hadamard gate.
Hadamard applied to zero is the state composed of half blue and half orange, which we denote by $(1/2,1/2)$. 
Similarly Hadamard applied to orange is $(1/2,-1/2)$.  
If we apply Hadamard on $(1/2,1/2)$, 
we apply it to both parts.
The blue  half becomes $(1/2, 1/2)$ and the  
orange half becomes $(1/2,-1/2)$, so we obtain four parts, which we will write $(1/4,1/4,1/4,-1/4)$.
Up to this stage, our representation accurately  reflects the effect of the gate (up to normalization).
What can get us into trouble is interference.

In some cases we can add the parts of same sign and let parts of opposite signs cancel.
In the case of Hadamard,
we are tempted to say that the orange parts with opposite signs cancel, 
leaving an all-blue disk.  In this case, this correctly mimics what happens 
to the amplitudes,
and it allows us to 
convey a message that is essentially correct:
the fact that applying Hadamard twice to basis states is identity, and this 
occurs because 
of interference.
It is also all that is needed to explain how and 
why some simple applications, such as the BB84 key exchange protocol, or Deutsch's algorithm, work.

In general, however, we refrain from adding together and rearranging the
slices in our representation. The first reason is that positive and negative interference does 
not usually behave well in our representation. But there is a second very imoirtant reason which 
is that 
when multiple qubits are entangled, moving parts around would affect the entanglement
between the qubits. We explain this is more detail in Section~\ref{sec:2qubitgates}.

We consider the effect of single qubit gates 
$G$, with $G\ket{0}=a\ket{0}+b\ket{1}$ and 
$G\ket{1}=c\ket{0}+d\ket{1}$ 
We will say that in our representation, $G(P,Q) = (A,B,C,D)$ where
$$ A = P\cdot   \sgn(a) |a|^2
\quad  B = P \cdot \sgn(b) |b|^2 
 \quad C = Q \cdot \sgn(c)|c|^2 
 \quad D = Q \cdot \sgn(d)|d|^2$$ 
In the schematic representation, we obtain four parts, starting from the top going clockwise with blue, and alternating blue and orange parts.

\begin{figure}[h]
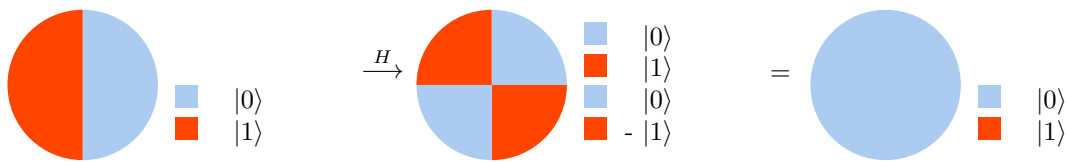

\centering
\qubobOne{50}{50}
		{ \parbox[b][1.5cm][t]{5mm}{$\stackrel{H}{\longrightarrow}$}} 
	\qubobOneSplit{25}{25}{25}{-25} 
		{ \parbox[b][1.2cm][t]{3mm}{=}}  
	\qubobOne{100}{0}
\caption{Applying Hadamard to $\frac{1}{\sqrt{2}}(\ket{0}+\ket{1})$ behaves well since the positive and negative parts cancel completely.
}
\end{figure}

\begin{figure}
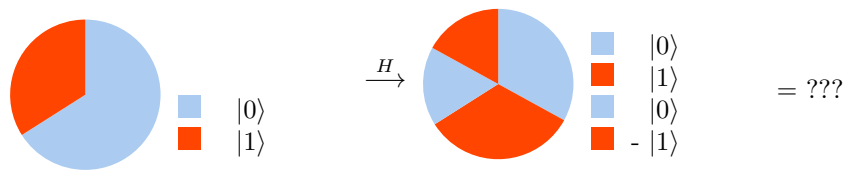

\centering
\qubobOne{66}{34}
		{ \parbox[b][1.5cm][t]{5mm}{$\stackrel{H}{\longrightarrow}$}} 
\qubobOneSplit {33}{33}{17}{-17}
		{ \parbox[b][1.2cm][t]{30mm}{= ???}}  

\caption{Hadamard applied to other states does not behave well. 
Applying Hadamard to $\frac{\sqrt{2}}{\sqrt{3}}\ket{0} + \frac {1}{\sqrt{3}}\ket{1}$, we get $\frac{1}{\sqrt{3}}\ket{0}+\frac{1}{\sqrt{3}}\ket{1} + \frac{{1}}{\sqrt{6}}\ket{0}-\frac{{1}}{\sqrt{6}}\ket{1}$.
Our representation suggests that there should be $\frac{1}{3}-\frac{1}{6}$ orange (1) left, whereas the
correct fraction is $(\frac{1}{\sqrt{3}}-\frac{1}{\sqrt{6}})^2$. 
}
\end{figure}

Henceforth we will use the notation $(A,B,C,D)$ to describe a disc with four areas, whose colors are blue, orange, blue, orange.
For example, if Hadamard is applied to $(1/2,1/2)$, we get $(1/4,1/4,1/4,-1/4)$. 
Unless the state is entangled with another, we start from the top with a blue slice of size $A$, and so forth, going clockwise.

\subsection{Applying gates on two qubits} 

\label{sec:2qubitgates}

Applying two-qubit gates works similarly to the one-qubit case.

Recall that two qubits in our representation, $(P,Q)$ and $(P',Q')$, 
do not fully determine the two-qubit state, since the angle at which they are
placed determines how they are correlated.
If the first qubit is blue going clockwise from the top, and the second disk is
orange starting from an angle of  $\theta\cdot 2\pi$,  going clockwise,
then (ignoring for the sake of simplicity any possible 
phases) the state represented is
$$ \ket{\psi} = \sqrt{\theta} \ket{00} + \sqrt{P{-}\theta}\ket{01} +
\sqrt{1{-}P{-}P'{+}\theta} \ket{11} + \sqrt{P'{-}\theta} \ket{10}.$$
We will use the notation $\ket{\psi} \mapsto (P,Q)\theta(P',Q')$.

When we apply a  single qubit gate to the first qubit of a two-qubit state, we apply it while  preserving the angle between the two qubits. To ensure that entanglement is preserved, we apply it to all four areas corresponding to the basis elements $\ket{00},\ket{01},\ket{11}, \ket{10}$.  These four areas are easily seen in the stacked representation. Each of the four areas will be further subdivided by gates such as Hadamard.

Similarly, when applying a two-qubit gate, we apply it to all four areas. Control gates are particularly easy to visualize. (See Figure~\ref{fig:CNOT}.) In the area where the control bit is 0, we do nothing, and in the area where the control bit is 1, we apply the gate as usual.

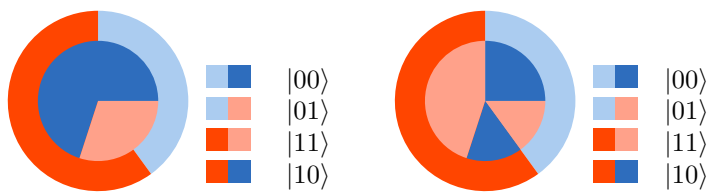
\begin{figure}[h]
\centering
\qubobsStacked{25}{15}{15}{45}
\begin{tikzpicture}
	\qubobPie{90}{0,0}{\outerdiscsize}{Qblue1,Qorange1}
	{ {40}/ , {60}/ }
	\qubobPie{90+0*3.6}{0,0}{\innerdiscsize}{Qblue2,Qorange2, Qblue2, Qorange2 }
	{25/ ,  15/ , 15/, 45/ }
\end{tikzpicture}
\newlegend{~,~,~,~}

\caption{The CNOT gate is applied to the qubit on the left. The control bit is  the outer bit.
When the outer bit is blue, the inner bit remains the same. When the outer bit is orange, the colors of the inner bit are flipped.}
\label{fig:CNOT}
\end{figure}

\section{Applications}
\subsection{Key exchange}

We have used our representation to explain the BB84 key agreement protocol~\cite{BB84}  to high school students
and members of the general public.
We distribute to participants a kit containing 
\begin{itemize}
\item Qubits with $\ket{0}$ (blue) on one side and $H\ket{0}$ on the other, 
\item Qubits with $\ket{1}$ (orange) on one side and $H\ket{1}$ on the other, 
\item Two dice, one to pick a bit, and one to apply, or not, Hadamard.
\item Eight numbered black envelopes in which qubits can be placed and observed on either side using small cutout flaps.
\item A grid to mark the results: what value was selected, what basis was used, and what measurement was made, and what the outcome was.
\end{itemize}

\begin{figure}[h]
\centering 
\includegraphics[height=3cm]{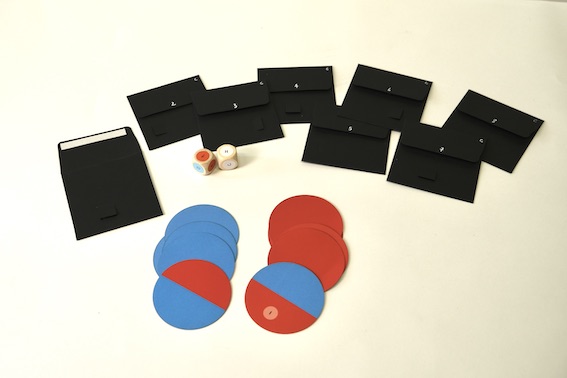}
\includegraphics[height=3cm]{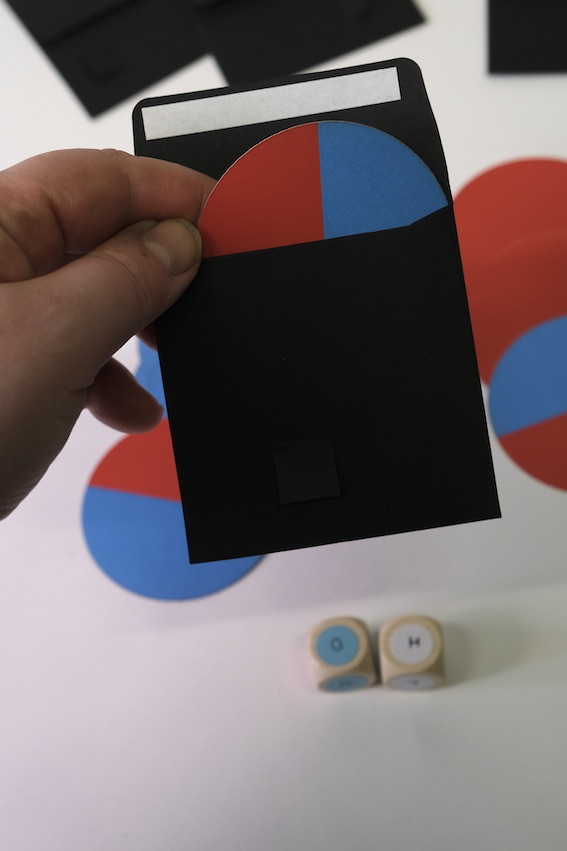}
\includegraphics[height=3cm]{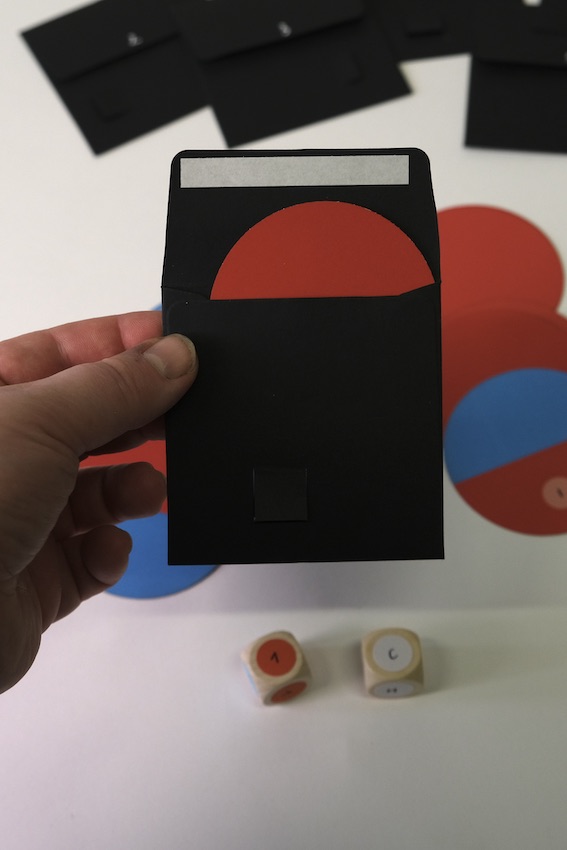}
\includegraphics[height=3cm]{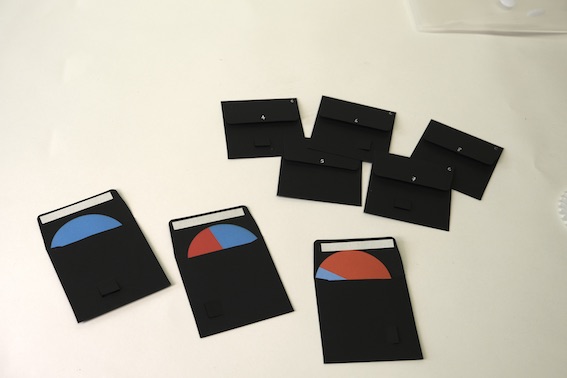}
\caption{Our BB84 kit comes with eight envelopes and two dice. Alice picks 8 qubits and places each them in a numbered envelope. Qubits can be measured by opening a small flap in the envelope.
Flipping a qubit over changes the basis. Flipping over the envelope changes the measurement basis.}
\end{figure}

Participants playing Alice can prepare qubits in the standard or diagonal basis by placing them in envelopes with the correct side facing up. 
A second participant playing Bob 
can measure a qubit in the standard basis by opening a flap on the front of the envelope, or in the Hadamard basis by turning over the envelope (and hence the qubit inside) and
opening a flap on the back of the envelope. All the results are marked on a grid.
Participants can see for themselves that any time the basis chosen for the qubit and the basis for the measurement coincide, the outcome of the measurement made by 
Bob equals the bit that was chosen by Alice.
We can then explain that any discrepancy would be due to an observation 
made by a third party.

\subsection{Teleportation}

We like to explain quantum teleportation~\cite{BBCJPW} in two steps. First we 
give a simplified and fully classical protocol to teleport a coin with an arbitraty bias. 
This protocol only uses a CNOT gate, and Alice measures only one of her qubits and sends the outcome
to Bob.
This is enough for Bob to obtain the bias of Alice's coin.

Splitting the protocol into two steps, a fully classical step with CNOT and the truly quantum part with Hadamard
demystifies how and why the protocol works, and allows us to clearly identify what part of the magic
is due to classical correlations and what part is attributable to quantum.

\subsubsection{Teleporting a biased coin}

At the start of the protocol, Alice has a biased coin, and Alice and Bob share one pair of 
perfectly correlated unbiased coins. This is similar to Alice having a qubit without any phase
and the two players sharing an EPR pair.

In Figure~\ref{fig:CNOT}, the first stacked pair of qubits belongs to Alice and
is composed, on the inner disk, of the qubit (biased coin) 
that Alice wishes to transmit, and on the outer disk, half of an EPR  pair (or correlated coins).
The rightmost qubit is Bob's half of the EPR pair. 

In Step 2 of the simplified protocol, Alice applies a CNOT operation on her qubits, 
using the inner disk as the control qubit.

In Step 3, Alice measures her outer qubit. Since the qubits are entangled (correlated), this also affects Bob's part of the EPR pair.
In Step 3a, if Alice measures 0 (blue) on the outer disk, Bob gets the correct qubit (without the phase).
In Step 3b, if Alice measures 1 (orange) on the outer disk, Bob needs to apply a correction 
to reverse the values.
\begin{figure}[h]
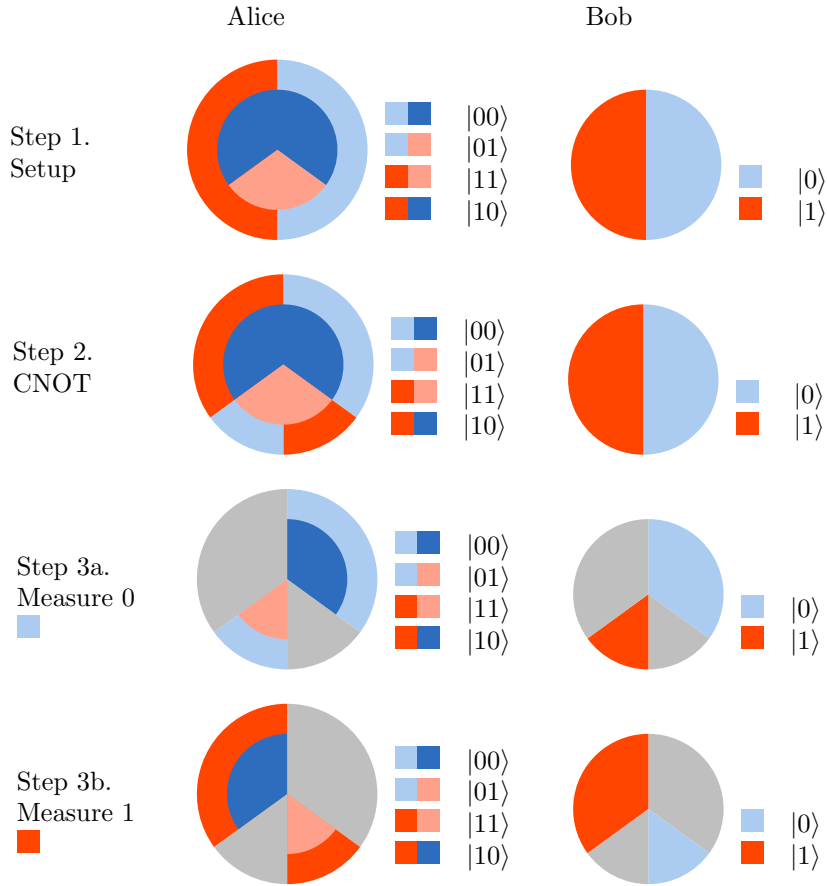

\hspace{5cm} Alice\hspace{4cm}Bob
\bigskip

\centering
 \parbox[b][1.5cm][t]{20mm}{ Step~1.\\ Setup}  
 \teleportationSetup{70}{30}

\bigskip

\parbox[b][1.5cm][t]{20mm}{Step~2. \\CNOT} 
\teleportationCNOT{70}{30}

\bigskip

\parbox[b][1.5cm][t]{20mm}{Step~3a. \\Measure 0 \\ \qbox{Qblue1}} 
\teleportationMeasureZero{70}{30}

\bigskip

\parbox[b][1.5cm][t]{20mm}{Step~3b. \\Measure 1 \\ \qbox{Qorange1}}
\teleportationMeasureOne{70}{30}

\caption{Classical teleportation of a qubit without a phase. Alice's two qubits (or coins) 
are represented on the left, and Bob's qubit (or coin) is on the right.}
\label{fig:CNOT}
\end{figure}

When the measurement outcome is 0 (blue), the residual state on Bob's side has the correct bias.
When the outcome is 1, the bias is correct provided he flips the colors.
Notice that we have given a fully classical protocol that transmits one coin flip with an arbitrary bias,
using a pair of correlated unbiased coins plus the transmission of an unbiased coin flip.

\subsubsection{Full teleportation protocol}

\begin{figure}[h]
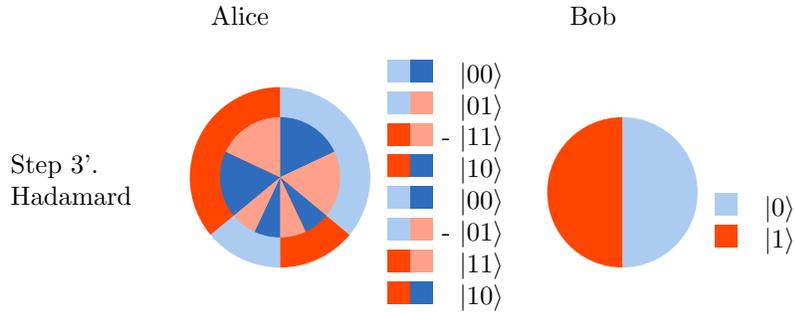

\hspace{5cm} Alice\hspace{4cm}Bob

\bigskip

\centering 
\parbox[b][1.5cm][t]{20mm}{Step~3'. \\Hadamard}
\teleportationH{72}{28}

\caption{Applying the Hadamard gate to the inner qubit before measuring.}
\label{fig:H}

\end{figure}

In order for Alice to transmit the phase of her qubit, she adds the truly quantum part of 
the protocol by applying Hadamard to her inner qubit. This will further split 
all the slices into two as we illustrate in Figure~\ref{fig:H}.

\begin{figure}[h]
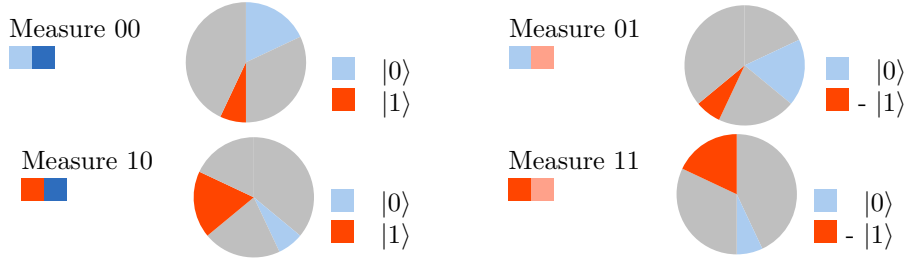

\centering
\parbox[b][1.5cm][t]{20mm}{Measure 00\\ \qbox{Qblue1}\qbox{Qblue2} }
\teleportationMeasureZZ{72}{28}
\parbox[b][1.5cm][t]{20mm}{Measure 01\\ \qbox{Qblue1}\qbox{Qorange2}}
\teleportationMeasureZO{72}{28}

\parbox[b][1.5cm][t]{20mm}{Measure 10\\ \qbox{Qorange1}\qbox{Qblue2}}
\teleportationMeasureOZ{72}{28}
\parbox[b][1.5cm][t]{20mm}{Measure 11\\ \qbox{Qorange1}\qbox{Qorange2}}
\teleportationMeasureOO{72}{28}

\caption{The residual state on Bob's side for the four possible measurement outcomes.}
\label{fig:M}
\end{figure}

The four possible measurements oucomes and corresponding residual state on Bob's side
are given in Figure~\ref{fig:M}.
Here we can see that in two of the cases (on top), the proportions are correct, and in the remaining two
cases (on the bottom), the colors need to be flipped. Similarly, we can see that in two of the cases (on the left),
the signs are correct and in the remaining two (on the right), the signs need to be flipped.

\pagebreak
\section{Conclusions}

We have presented new physical devices and a visual representation of qubits 
that can be used for teaching or for outreach activities. 
We have found it particularly helpful in conveying concepts such as entanglement
and interference, concepts that key to understanding quantum algorithms, but can be
difficult to grasp at an intuitive level, especially for audiences who are
not accustomed to using mathematical formalism. 

We have used our devices and representation in the classroom in 
addition to the usual mathematical notation and
have found that it lends itself particularly well to
walking through quantum circuits to see how and why they work.

Our representation has its drawbacks, the main one being that in general,
applying linearity in the usual way on the squared-amplitudes vector does 
not yield 
the same result as on the amplitudes. However, 
by being cautious with interference, 
and checking whether a particular 
operation behaves well on the states that they are being applied to,
it does work in sufficiently many cases so as to allow us to
explain some basic algorithms and protocols.

In forthcoming work, we plan to extend our applications to include more algorithms 
as well as nonlocality with the CHSH game. 
Also in the works is an interactive interface and a musical representation of
qubits so that it can be more widely accessible.  
More information on our project can be found on the website
\url{https://qubobs.irif.fr}.

\section*{Acknowledgements}
We would like to thank the students and participants in our 
outreach activities for their feedback. 
We thank Maris Ozols for references to tensor diagrams and Serge Massar for
references to some of the other representations.
We are grateful to the many colleagues who served as sounding boards 
throughout the different versions of this work. 

This project was funded by IRIF, by a grant from 
IDEX Universit\'e Paris Cit\'e, and the ANR grant 
SAPS-RA-MCS QUBOBS.

\bibliography{qubobs}

\end{document}